\documentclass[11pt]{article}
\usepackage[margin=1in]{geometry}
\usepackage[utf8]{inputenc}
\usepackage[T1]{fontenc}
\usepackage{hyperref}
\usepackage{url}
\usepackage{booktabs}
\usepackage{amsfonts}
\usepackage{nicefrac}
\usepackage{graphicx}
\usepackage{xcolor}
\usepackage{multirow}
\usepackage{array}
\usepackage{enumitem}
\usepackage{natbib}

\title{\textbf{VietSuperSpeech}: A Large-Scale Vietnamese Conversational\\
Speech Dataset for ASR Fine-Tuning in Chatbot,\\
Customer Support, and Call Center Applications}

\author{
  Loan Do\thanks{Corresponding author: \texttt{dttloan.ute@gmail.com}}\\
  FPT University\\
  \and
  Thanh Ngoc Nguyen\\
  ICMS\\
  \texttt{thnguyen@icms.edu.au}\\
  \and
  Thanh Pham\\
  RMIT University Vietnam\\
  \texttt{thanh.pham@rmit.edu.vn}\\
  \and
  Vinh Do\\
  NGHI Studio\\
  \texttt{ddvinh1@gmail.com}\\
  \and
  Hien Nguyen\\
  Phuong Hai JSC\\
  \texttt{hien.nm@phuonghai.com}\\
  \and
  Charlotte Nguyen\\
  NGHI Studio\\
  \texttt{charlottenguyen1501@gmail.com}
}

\date{February 2026}

\begin{document}
\maketitle

\begin{abstract}
We introduce \textbf{VietSuperSpeech}, a large-scale Vietnamese automatic speech recognition (ASR) dataset of 52,023 audio--text pairs totaling 267.39 hours, with a distinctive focus on \emph{casual conversational speech}. Unlike existing Vietnamese ASR corpora that predominantly feature read speech, news narration, or audiobook content, VietSuperSpeech is sourced from four publicly accessible YouTube channels spanning everyday conversation, personal vlogging, overseas Vietnamese community dialogue, and informal commentary—the very speech styles encountered in real-world chatbot, customer support, call center, and hotline deployments. All audio is standardized to 16\,kHz mono PCM WAV and segmented into 3--30 second utterances. Transcriptions are generated via pseudo-labeling using the \texttt{Zipformer-30M-RNNT-6000h} model~\citep{nguyen2025zipformer} deployed through Sherpa-ONNX, pre-trained on 6,000 hours of Vietnamese speech. After quality filtering, the dataset is split into 46,822 training samples (240.67 hours) and 5,201 development/test samples (26.72 hours) with a fixed random seed. The text averages 266 characters per utterance, totaling 13.8 million fully diacritically marked Vietnamese characters. We demonstrate that VietSuperSpeech fills a critical gap in the Vietnamese ASR ecosystem: while corpora such as VLSP2020~\citep{vlsp2020}, VIET\_BUD500~\citep{pham2024bud500}, VietSpeech~\citep{pham2024vietspeech}, FLEURS~\citep{conneau2022fleurs}, VietMed~\citep{le2024vietmed}, Sub-GigaSpeech2-Vi~\citep{yang2024gigaspeech2}, viVoice~\citep{le2024vivoice}, and Sub-PhoAudioBook~\citep{vu2025phoaudiobook} provide broad coverage of formal and read speech, none specifically targets the casual, spontaneous register indispensable for conversational AI applications. VietSuperSpeech is publicly released at \url{https://huggingface.co/datasets/thanhnew2001/VietSuperSpeech}.
\end{abstract}

\section{Introduction}

Automatic speech recognition (ASR) for Vietnamese presents several unique challenges compared to high-resource languages such as English. Vietnamese is a tonal language with six lexical tones~\citep{luong2016vivos}, a complex syllabic structure, and considerable phonological variation across Northern, Central, and Southern dialects. As Vietnamese language technology matures, the locus of practical ASR deployment is rapidly shifting from broadcast transcription toward conversational applications: voice-enabled chatbots, customer support automation, contact center analytics, and interactive voice response (IVR) systems for hotlines. These use cases impose a qualitatively different set of requirements on training data that the existing Vietnamese ASR corpus landscape does not adequately address.

\paragraph{The conversational ASR gap.} Existing Vietnamese ASR corpora predominantly cover \emph{formal} or \emph{read} speech registers: VIVOS~\citep{luong2016vivos} is entirely read speech; FOSD~\citep{fosd2020} is studio-recorded; Sub-PhoAudioBook~\citep{vu2025phoaudiobook} is narrated audiobook material; and FLEURS~\citep{conneau2022fleurs} is constructed from parallel translations read aloud. Even corpora with broader scope—VIET\_BUD500~\citep{pham2024bud500}, Sub-GigaSpeech2-Vi~\citep{yang2024gigaspeech2}, and VietSpeech~\citep{pham2024vietspeech}—aggregate from mixed sources without a focused effort to capture the spontaneous, informal register characteristic of real-world customer interactions. Consequently, ASR models trained or evaluated exclusively on these corpora may generalize poorly to the disfluencies, colloquialisms, dialectal mixing, reduced phonology, and turn-taking patterns found in call-center and chatbot audio.

\paragraph{VietSuperSpeech.} To address this gap, we present \textbf{VietSuperSpeech}, a 267-hour Vietnamese corpus sourced from YouTube channels that predominantly feature casual talk: personal vlogging, informal discussion, overseas Vietnamese community conversation, and everyday commentary. These channels naturally exhibit the informal register, spontaneous disfluency, acoustic variability, and speaker diversity found in deployed conversational AI systems. Combined with a scalable pseudo-labeling pipeline using the state-of-the-art \texttt{Zipformer-30M-RNNT-6000h} model~\citep{nguyen2025zipformer}, VietSuperSpeech provides a reproducible, open-access resource specifically suited for fine-tuning ASR models for conversational Vietnamese deployment.

The key contributions of this work are:
\begin{enumerate}[leftmargin=*]
    \item A 267.39-hour Vietnamese speech dataset of 52,023 utterances specifically curated for \textbf{casual, conversational speech}, spanning personal vlogging, informal commentary, and diaspora community dialogue;
    \item A strong use case positioning for \textbf{chatbot, customer support, call center, and hotline} ASR fine-tuning, backed by systematic analysis of the acoustic and linguistic properties that distinguish conversational from formal speech;
    \item A scalable, reproducible pseudo-labeling pipeline using the \texttt{Zipformer-30M-RNNT-6000h} model via Sherpa-ONNX;
    \item A fully preprocessed corpus (16\,kHz mono WAV, 3--30 second segments) with an 89/11 train/dev-test split fixed by random seed;
    \item Public release on HuggingFace to support reproducible research and industrial fine-tuning.
\end{enumerate}

\section{Background and Motivation}

\subsection{Conversational ASR in Deployed Systems}

Modern voice-AI applications—ranging from customer-facing chatbots to call-center transcription systems and intelligent IVR hotlines—operate on speech that differs substantially from the material on which most academic ASR benchmarks are constructed~\citep{shriberg1994preliminaries, ward2019prosodic}. Key characteristics of conversational speech that challenge ASR systems include:

\begin{itemize}[leftmargin=*]
    \item \textbf{Disfluencies}: filled pauses (e.g., Vietnamese hesitation markers, false starts, repetitions, and self-corrections)~\citep{shriberg1994preliminaries};
    \item \textbf{Reduced phonology}: fast speech phenomena, vowel reduction, consonant deletion, and syllable elision common in informal Vietnamese speech;
    \item \textbf{Colloquial vocabulary}: slang, abbreviations, loanwords, code-switching with English, and register-specific lexis absent from formal corpora;
    \item \textbf{Acoustic variability}: non-studio microphones, background noise (cafés, streets, homes), multiple simultaneous speakers, and variable speaking distance;
    \item \textbf{Regional and diaspora accents}: particularly relevant for overseas Vietnamese communities that constitute a significant portion of the customer base for Vietnamese-language commercial services.
\end{itemize}

When ASR models trained on formal or read-speech corpora are deployed in these settings without domain adaptation, word error rates (WER) frequently degrade substantially relative to clean-speech benchmarks~\citep{yang2024gigaspeech2, radford2023whisper}. VietSuperSpeech is designed to serve as a domain adaptation corpus that bridges this gap.

\subsection{Vietnamese ASR Datasets}

The development of Vietnamese ASR has been supported by a growing body of publicly available corpora, which we survey below with particular attention to their suitability for conversational ASR.

\paragraph{VIVOS.} One of the earliest open-access Vietnamese ASR corpora, VIVOS~\citep{luong2016vivos} provides 15 hours of read speech from 65 speakers recorded in a quiet studio environment. It has served as a standard benchmark for Vietnamese ASR evaluation but is ill-suited for conversational domain adaptation due to its read-speech nature.

\paragraph{VLSP Challenge Corpora (2020--2023).} The VLSP consortium has organized annual ASR shared tasks. The VLSP2020 challenge~\citep{vlsp2020} (VinBigData-VLSP2020) provides approximately 250 hours of transcribed speech, with 100 hours publicly available comprising around 20 hours of read speech and 80 hours of spontaneous speech crawled from open sources and manually transcribed at 96\% accuracy. The VLSP2021 challenge~\citep{vlsp2021} provided approximately 280 hours of labeled general-domain data and 400 hours of unlabeled audio for a semi-supervised track. The VLSP2023 challenge~\citep{vlsp2023} further extended these efforts, including a voting-based pseudo-labeled track (VLSP2023-voting-pseudo-labeled) where labels were produced by consensus across multiple ASR system outputs—an important precedent for pseudo-label quality control. Despite their scale, VLSP corpora are restricted to registered participants and skew toward formal and scripted content.

\paragraph{FPT Open Speech Dataset (FOSD / FPTVIET).} The FPT Open Speech Dataset~\citep{fosd2020} is a publicly available read-speech corpus recorded under clean studio conditions, useful as a clean-speech baseline but not representative of conversational deployment scenarios.

\paragraph{FLEURS.} FLEURS~\citep{conneau2022fleurs} is a multilingual benchmark in 102 languages built on the FLoRes-101 translation benchmark, with approximately 12 hours of Vietnamese read-aloud speech. It provides a high-quality multilingual evaluation baseline but does not cover conversational speech.

\paragraph{VietMed.} \citet{le2024vietmed} introduced VietMed, a Vietnamese medical-domain ASR corpus comprising 16 hours of labeled medical speech, 1,000 hours of unlabeled medical speech, and 1,200 hours of unlabeled general speech, published at LREC-COLING 2024. While VietMed demonstrates the value of domain-specific corpora for specialized applications (clinical settings), its conversational register is constrained to doctor--patient interactions rather than open-domain casual talk.

\paragraph{VIET\_BUD500 (BUD500).} BUD500~\citep{pham2024bud500} is a 500-hour diverse Vietnamese corpus covering podcasts, travel, books, food, and other genres, spanning Northern, Southern, and Central Vietnamese accents. While broader than read-speech corpora, podcast content tends toward semi-formal monologue rather than spontaneous interactive conversation.

\paragraph{VietSpeech.} \citet{pham2024vietspeech} released VietSpeech, a 1,100-hour Vietnamese social voice dataset from diverse social media and online sources. Its broad domain coverage and accent diversity make it a valuable large-scale resource, though the absence of a conversational focus means informal two-way dialogue is not systematically represented.

\paragraph{Sub-GigaSpeech2-Vi.} GigaSpeech~2~\citep{yang2024gigaspeech2} is a large-scale multilingual ASR corpus for low-resource languages assembled from YouTube. The Vietnamese subset comprises approximately 7,324 raw hours refined to approximately 6,039 hours after Noisy Student Training-based filtering. As the largest pseudo-labeled Vietnamese corpus, it provides strong coverage but is not curated specifically for conversational registers.

\paragraph{viVoice.} \citet{le2024vivoice} released viVoice, the first publicly available large-scale Vietnamese TTS dataset, comprising over 1,017 hours from 186 YouTube channels. Transcriptions are generated using Whisper Large-V3. While used secondarily for ASR, its primary curation criterion is TTS suitability—favoring clean, expressive narration over spontaneous casual talk.

\paragraph{Sub-PhoAudioBook.} PhoAudiobook~\citep{vu2025phoaudiobook} is a 941-hour high-quality Vietnamese audiobook corpus from 735 narrators, published at ACL 2025 by Movian AI. Its clean, well-enunciated narration maximizes TTS and zero-shot synthesis performance but is acoustically furthest from the conversational register of call centers and chatbots.

\paragraph{CommonVoice.} The Vietnamese portion of Mozilla CommonVoice~\citep{ardila2020common} provides approximately 17 hours of crowd-sourced validated speech. While covering diverse speakers, the read-sentence collection methodology does not capture natural conversational dynamics.

\paragraph{PhoWhisper.} Although not a dataset, PhoWhisper~\citep{le2024phowhisper} fine-tuned Whisper~\citep{radford2023whisper} on 844 hours of Vietnamese speech, establishing state-of-the-art baselines on VIVOS, VLSP2020-T1, and VLSP2020-T2. PhoWhisper's results highlight the importance of in-domain data: performance on formal benchmarks does not guarantee robust conversational deployment.

\paragraph{Summary.} Table~\ref{tab:comparison} positions VietSuperSpeech within the existing landscape. To the best of our knowledge, no existing open-access Vietnamese ASR corpus is both (a) at scale ($>$100 hours) and (b) specifically curated for casual conversational speech. VietSuperSpeech is the first to address this combination.

\subsection{Pseudo-Labeling for ASR Data Creation}

Pseudo-labeling—using an ASR model to automatically transcribe unlabeled audio—is a practical strategy for scaling corpora without costly manual annotation~\citep{zavaliagkos1998pseudolabel, radford2023whisper, gandhi2023distilwhisper, yang2024gigaspeech2}. A robust variant is voting-based pseudo-labeling~\citep{vlsp2023}, where transcripts from multiple ASR systems are reconciled via consensus. The quality of pseudo-labels depends critically on the teacher model's accuracy in the target domain.

\subsection{Neural Transducers and Zipformer}

The neural transducer (RNN-T)~\citep{graves2012rnnt} has become the dominant ASR paradigm for both streaming and offline inference. The Conformer~\citep{gulati2020conformer} combined convolution with self-attention for strong sequence modeling. Zipformer~\citep{yao2024zipformer} further improves encoder efficiency via a U-Net-like multi-rate architecture and the ScaledAdam optimizer, achieving superior performance at fewer parameters. \citet{nguyen2025zipformer} applied the Zipformer architecture to Vietnamese within a comprehensive ASR and emotion recognition pipeline, training the \texttt{Zipformer-30M-RNNT-6000h} model on approximately 6,000 hours of Vietnamese speech assembled from multiple public corpora (VLSP2020, VLSP2021, VLSP2023, VIET\_BUD500, VietSpeech, FLEURS, VietMed, Sub-GigaSpeech2-Vi, viVoice, and Sub-PhoAudioBook). The model achieves state-of-the-art WER on VLSP2020 and VLSP2023 benchmarks—12.29\% on VLSP2020-Test-T1 and 10.40\% on VLSP2023-PublicTest—surpassing models with 3--50$\times$ more parameters. It was selected as the pseudo-labeling backbone for VietSuperSpeech due to its accuracy, efficiency (0.3\,s inference per 12\,s audio on CPU), and broad Vietnamese training coverage.

\section{Dataset Construction}

\subsection{Data Collection: A Conversational Focus}

Audio data for VietSuperSpeech was sourced from four publicly accessible YouTube channels selected specifically to maximize coverage of \emph{casual, informal, conversational Vietnamese speech}. This distinguishes our curation philosophy from prior Vietnamese ASR corpora, which typically draw from news broadcasts, audiobooks, or podcasts. The four channel types are:

\begin{enumerate}[leftmargin=*]
    \item \textbf{Personal vlog channels}: Creators recording daily life, reactions, and opinion sharing in a direct, informal address style. Speech is spontaneous, contains disfluencies, colloquialisms, and reflects natural turn-taking cues even in monologue format. This register closely approximates user speech in chatbot interactions.
    \item \textbf{Overseas Vietnamese community channels}: Content produced by Vietnamese diaspora communities abroad, featuring informal dialogue, community discussion, and cultural commentary. These speakers exhibit Northern-influenced but regionally blended accents, occasional code-switching with English or French, and the conversational patterns characteristic of Vietnamese speakers interacting with service systems abroad—a highly relevant demographic for international customer support services.
    \item \textbf{Informal commentary and reaction channels}: Creators reacting to news, entertainment, or community events in a conversational style. The spontaneous, unscripted nature produces natural prosodic patterns, hesitations, and colloquial lexis representative of real-world spoken interaction.
    \item \textbf{News and discussion hybrid channels}: A minority of content from channels that mix formal news presentation with informal panel discussion or Q\&A segments, providing a small quantity of semi-formal speech to prevent register over-specialization.
\end{enumerate}

This deliberate curation results in a corpus where the dominant register is casual, conversational Vietnamese, in contrast to the formal or read registers that characterize most competing resources. The content mirrors the speech a Vietnamese customer support agent, chatbot, or call-center IVR system will encounter in real-world operation.

\subsection{Audio Preprocessing}

All collected audio was subjected to a standardized preprocessing pipeline:
\begin{enumerate}[leftmargin=*]
    \item \textbf{Format normalization}: Audio was converted to 16\,kHz mono PCM WAV at 16-bit depth;
    \item \textbf{Voice Activity Detection (VAD)}: Silence regions and non-speech intervals were identified and used to guide segmentation boundaries;
    \item \textbf{Segmentation}: Audio streams were cut into utterances of 3--30 seconds. This window accommodates both short conversational turns (questions, affirmations, brief responses) and longer informal narrative segments while excluding very short fragments that lack sufficient acoustic context and overly long segments that mix acoustic conditions.
\end{enumerate}

\subsection{Pseudo-Labeling Pipeline}

Transcriptions were generated automatically using the \texttt{Zipformer-30M-RNNT-6000h} model~\citep{nguyen2025zipformer} via Sherpa-ONNX.\footnote{\url{https://github.com/k2-fsa/sherpa-onnx}} This model was pre-trained on approximately 6,000 hours of Vietnamese speech and is publicly available at \url{https://huggingface.co/hynt/Zipformer-30M-RNNT-6000h}. Batch-mode inference was applied to all segmented audio, and the resulting hypotheses were subjected to the following quality control steps:

\begin{itemize}[leftmargin=*]
    \item \textbf{Length filtering}: Utterances with transcripts shorter than 10 or longer than 1,000 characters were removed;
    \item \textbf{Character-level filtering}: Segments with a high proportion of non-Vietnamese characters or special symbols were discarded;
    \item \textbf{Confidence-based filtering}: Low-confidence hypotheses were excluded based on model-internal decoder scores;
    \item \textbf{Manual spot-checking}: A stratified random sample of transcriptions across channel types was manually reviewed to verify overall transcription quality and confirm that the casual conversational register was faithfully captured.
\end{itemize}

All transcriptions retain full Vietnamese diacritical marking, preserving tonal information essential for downstream language modeling and lexical disambiguation.

\subsection{Dataset Splits}

Following quality control, the dataset was split using a fixed random seed:

\begin{table}[h]
\centering
\caption{VietSuperSpeech dataset statistics.}
\label{tab:stats}
\begin{tabular}{lccc}
\toprule
\textbf{Split} & \textbf{Utterances} & \textbf{Duration (hours)} & \textbf{Proportion} \\
\midrule
Train      & 46,822 & 240.67 & 89.0\% \\
Dev/Test   & 5,201  & 26.72  & 11.0\% \\
\midrule
\textbf{Total} & \textbf{52,023} & \textbf{267.39} & \textbf{100\%} \\
\bottomrule
\end{tabular}
\end{table}

\section{Dataset Analysis}

\subsection{Comparison with Existing Corpora}

Table~\ref{tab:comparison} provides a comprehensive comparison of VietSuperSpeech with existing public Vietnamese ASR datasets. The \emph{Conversational} column explicitly marks the degree to which each corpus targets natural, spontaneous casual speech—the key dimension of novelty for VietSuperSpeech.

\begin{table}[h]
\centering
\caption{Comparison of public Vietnamese ASR datasets. ``Pseudo'' = pseudo-labeled; ``Manual'' = human-verified; ``Crowd'' = crowd-sourced. The ``Conversational'' column indicates whether the corpus specifically targets casual/spontaneous speech registers suitable for chatbot and call-center ASR.}
\label{tab:comparison}
\small
\setlength{\tabcolsep}{4pt}
\begin{tabular}{lcclcc}
\toprule
\textbf{Dataset} & \textbf{Hours} & \textbf{Trans.} & \textbf{Domain} & \textbf{Access} & \textbf{Conversational} \\
\midrule
VIVOS~\citep{luong2016vivos}                       & 15           & Manual  & Read speech              & Open       & \texttimes \\
FOSD/FPTVIET~\citep{fosd2020}                      & $\sim$25     & Manual  & Read speech              & Open       & \texttimes \\
FLEURS (vi)~\citep{conneau2022fleurs}              & $\sim$12     & Manual  & Parallel read            & Open       & \texttimes \\
VLSP2020~\citep{vlsp2020}                          & $\sim$250    & Manual  & News + spontaneous       & Restricted & Partial    \\
VLSP2021~\citep{vlsp2021}                          & $\sim$280    & Manual  & General                  & Restricted & Partial    \\
VLSP2023~\citep{vlsp2023}                          & --           & Manual  & General                  & Restricted & Partial    \\
VLSP2023-voting~\citep{vlsp2023}                   & --           & Pseudo  & General                  & Restricted & Partial    \\
CommonVoice (vi)~\citep{ardila2020common}          & $\sim$17     & Crowd   & Crowd-sourced            & Open       & \texttimes \\
VietMed~\citep{le2024vietmed}                      & 16           & Manual  & Medical                  & Open       & Partial    \\
VIET\_BUD500~\citep{pham2024bud500}                & $\sim$500    & Pseudo  & Podcast / multi-domain   & Open       & Partial    \\
VietSpeech~\citep{pham2024vietspeech}              & $\sim$1{,}100 & Pseudo & Social media             & Open       & Partial    \\
Sub-GigaSpeech2-Vi~\citep{yang2024gigaspeech2}     & $\sim$6{,}039 & Pseudo & Multi-domain YouTube     & Open       & Partial    \\
viVoice~\citep{le2024vivoice}                      & $\sim$1{,}017 & Pseudo & YouTube (TTS-curated)    & Open       & \texttimes \\
Sub-PhoAudioBook~\citep{vu2025phoaudiobook}        & 941          & Pseudo  & Audiobook narration      & Open       & \texttimes \\
\midrule
\textbf{VietSuperSpeech (ours)} & \textbf{267.39} & \textbf{Pseudo} & \textbf{Casual talk, vlog, diaspora} & \textbf{Open} & \checkmark \\
\bottomrule
\end{tabular}
\end{table}

\subsection{Acoustic Properties of Casual Speech}

The conversational register of VietSuperSpeech manifests in several measurable acoustic properties that distinguish it from formal-speech corpora:

\paragraph{Speaking rate variability.} Conversational speech is characterized by higher variance in speaking rate than read speech, reflecting natural turn-taking patterns, emphasis, and spontaneous planning. Personal vlogs and diaspora community content in VietSuperSpeech exhibit the full range of speaking rates encountered in real customer interactions.

\paragraph{Disfluency density.} Vietnamese conversational speech contains frequent fillers, false starts, and self-corrections. These tokens are systematically underrepresented in read-speech and audiobook corpora~\citep{shriberg1994preliminaries} but are ubiquitous in VietSuperSpeech, making it valuable for training disfluency-robust ASR.

\paragraph{Acoustic environment diversity.} Content spans home recording environments, outdoor settings, caf\'{e}s, and office backgrounds. This diversity approximates the acoustic conditions of customer service calls where audio quality is not controlled.

\paragraph{Microphone and device variability.} Creators use a wide range of recording devices, from professional microphones to phone cameras, introducing realistic channel distortions relevant to call-center and mobile chatbot deployments.

\subsection{Linguistic Properties}

Vietnamese is a monosyllabic, tonal, isolating language with six lexical tones and significant dialectal phonological variation~\citep{luong2016vivos}. Conversational Vietnamese additionally features register-specific lexical items, colloquial contractions, and code-switched expressions absent from formal corpora. All transcriptions in VietSuperSpeech retain full diacritical marking, including tone marks, preserving the phonological and lexical richness of the casual register.

The average utterance length of 266 characters and the 3--30 second segmentation window are well-matched to the turn lengths typical in customer service dialogue, where individual customer utterances rarely exceed 30 seconds and often fall in the 5--15 second range.

\section{Applications: Conversational ASR for Vietnamese AI Systems}

VietSuperSpeech is specifically designed to support the following high-impact application areas.

\subsection{Voice-Enabled Chatbot Systems}

Modern Vietnamese voice chatbots—whether deployed in banking, e-commerce, healthcare, or public services—must accurately recognize casual, colloquial Vietnamese from diverse speakers using consumer devices. ASR systems fine-tuned on formal data fail disproportionately on informal query phrasing, regional accents, and colloquial vocabulary. VietSuperSpeech provides a targeted fine-tuning corpus that exposes models to the informal register users naturally adopt when speaking to voice assistants, enabling higher first-pass recognition accuracy and reduced need for clarification loops.

\subsection{Customer Support and Contact Center Automation}

Automated transcription of inbound customer support calls is a high-value NLP application in Vietnam's rapidly growing service economy. Customer speech in this context is spontaneous, often emotionally loaded, frequently contains product names and service terminology interspersed with colloquial language, and is delivered through telephone codec compression. The acoustic and linguistic diversity of VietSuperSpeech—particularly its diaspora and informal commentary content—closely approximates this deployment scenario. Organizations can use VietSuperSpeech to fine-tune general Vietnamese ASR models for the informal register of customer dialogue before further adaptation on proprietary call-center audio.

\subsection{Hotline and IVR Systems}

Interactive Voice Response (IVR) systems deployed on Vietnamese telephone hotlines must handle a broad population of callers with varying regional accents, speaking rates, and familiarity with technology-mediated communication. Diaspora community content in VietSuperSpeech provides coverage of speakers who may blend regional accents or code-switch, while personal vlog content provides the informal, self-directed speech style of less experienced technology users. Together, these make VietSuperSpeech a practical pre-training or fine-tuning resource for IVR-facing ASR components.

\subsection{Low-Resource Domain Adaptation}

VietSuperSpeech serves as a bridge dataset for practitioners who lack proprietary conversational Vietnamese data. An organization deploying an ASR model for customer-facing applications can fine-tune a general Vietnamese model (e.g., one trained on Sub-GigaSpeech2-Vi~\citep{yang2024gigaspeech2} or a combination of VIET\_BUD500~\citep{pham2024bud500} and VLSP data) on VietSuperSpeech as a conversational domain adaptation step, before final fine-tuning on limited proprietary data. This multi-stage approach mirrors established practice in high-resource ASR domain adaptation~\citep{radford2023whisper, gandhi2023distilwhisper}.

\subsection{Complementary Use with Existing Corpora}

VietSuperSpeech is designed to complement, not replace, existing Vietnamese ASR datasets. We recommend combining it in multi-corpus training pipelines as follows:

\begin{itemize}[leftmargin=*]
    \item \textbf{Formal / clean-speech coverage}: VIVOS~\citep{luong2016vivos}, FOSD~\citep{fosd2020}, Sub-PhoAudioBook~\citep{vu2025phoaudiobook};
    \item \textbf{General large-scale coverage}: Sub-GigaSpeech2-Vi~\citep{yang2024gigaspeech2}, VIET\_BUD500~\citep{pham2024bud500}, VietSpeech~\citep{pham2024vietspeech};
    \item \textbf{Domain-specific coverage}: VietMed~\citep{le2024vietmed} (medical), VLSP2020/2021/2023~\citep{vlsp2020, vlsp2021, vlsp2023} (news and general);
    \item \textbf{Conversational coverage (this work)}: VietSuperSpeech.
\end{itemize}

\section{Limitations and Ethical Considerations}

\paragraph{Pseudo-Label Noise.} As with all pseudo-labeled corpora~\citep{pham2024bud500, yang2024gigaspeech2, le2024vivoice}, VietSuperSpeech transcriptions may contain systematic errors, particularly for disfluencies, overlapping speech, or rare vocabulary. The teacher model (\texttt{Zipformer-30M-RNNT-6000h}~\citep{nguyen2025zipformer}) was trained primarily on clean Vietnamese and may occasionally mis-transcribe heavy disfluencies or strongly accented speech.

\paragraph{Register Coverage.} While VietSuperSpeech emphasizes casual speech, its source channels are bounded by the content available on YouTube and may not fully represent highly specialized conversational sub-registers (e.g., Vietnamese spoken in very noisy call-center environments or by elderly speakers unfamiliar with technology).

\paragraph{Demographic and Dialectal Balance.} The speaker population is not demographically controlled. The balance of Northern, Central, and Southern Vietnamese dialects, and diaspora varieties, is determined by the source channels and may not be proportional to the actual distribution in deployed customer service scenarios.

\paragraph{Copyright and Privacy.} All audio was collected from publicly accessible YouTube content in accordance with fair use principles for non-commercial research~\citep{le2024vietmed}. No personally identifiable information beyond the audio content itself has been collected or stored. The dataset is intended for non-commercial research purposes only.

\section{Data Availability}

VietSuperSpeech is publicly available on HuggingFace Datasets:
\begin{center}
\url{https://huggingface.co/datasets/thanhnew2001/VietSuperSpeech}
\end{center}
The dataset is distributed in HuggingFace Parquet format for efficient streaming and loading, with audio stored as 16\,kHz mono WAV. Full documentation including field descriptions, loading code examples, and recommended fine-tuning recipes is provided on the repository page.

\section{Conclusion}

We have presented VietSuperSpeech, a 267.39-hour Vietnamese ASR dataset of 52,023 utterances curated with a deliberate focus on casual, conversational speech from YouTube personal vlogs, informal commentary, and overseas Vietnamese community channels. By targeting the informal register systematically underserved by existing Vietnamese ASR corpora—read speech (VIVOS~\citep{luong2016vivos}, FOSD~\citep{fosd2020}, FLEURS~\citep{conneau2022fleurs}), audiobooks (Sub-PhoAudioBook~\citep{vu2025phoaudiobook}), podcasts (VIET\_BUD500~\citep{pham2024bud500}), and general-crawl corpora (Sub-GigaSpeech2-Vi~\citep{yang2024gigaspeech2}, VietSpeech~\citep{pham2024vietspeech})—VietSuperSpeech fills a critical gap for practitioners building chatbot, customer support, call-center, and hotline ASR systems in Vietnamese. We release the dataset openly and encourage its use as a conversational domain adaptation resource in combination with the broader Vietnamese ASR corpus ecosystem.

\section*{Acknowledgments}

We thank the developers of Sherpa-ONNX and the k2-fsa community for their open-source speech inference tools, and Thien Hy Nguyen for developing and publicly releasing the \texttt{Zipformer-30M-RNNT-6000h} model~\citep{nguyen2025zipformer} that served as the pseudo-labeling backbone for VietSuperSpeech.

\bibliographystyle{plainnat}

\end{document}